\documentclass[11pt,amstex,a4paper]{article}
\usepackage[dvips]{graphicx}
\usepackage{amsmath}\usepackage{times}
\usepackage{psfig}
\usepackage{french}
\usepackage{epsfig}
\usepackage{amssymb}
\usepackage{theorem}

\newcommand{\dd}{\text{d}}
\newcommand{\sR}{\text{\tiny R}}
\newcommand{\ee}{\text{e}}

\newcommand{\p}{\partial}

\newcommand{\eps}{\varepsilon}

\numberwithin{equation}{section}
\begin{document}
\thispagestyle{empty}
\title{\bf Fermionic field theory for\\directed percolation in (1+1)-dimensions} 
\author{Vivien Brunel$^a$, Klaus Oerding$^b$ and Fr\'ed\'eric van Wijland$^c$\\\\
$^a$Service de Physique Th\'eorique\\
CEA Saclay\\
91191 Gif-sur-Yvette cedex, France\\\\
$^b$Institut f\"ur Theoretische Physik III\\
Heinrich Heine Universit\"at\\
40225 D\"usseldorf, Germany\\\\
$^c$Laboratoire de Physique Th\'eorique$^1$\\
Universit\'e de Paris-Sud\\
91405 Orsay cedex, France}

\maketitle
\vspace{-1cm}
\begin{small}
\begin{center}{\bf Abstract}\end{center}
{\small We formulate directed percolation in $(1+1)$ dimensions in the language of
a reaction-diffusion process with
exclusion taking place in one space dimension. We map
the master equation that describes the dynamics of the system onto a quantum
spin chain problem. From there we build an interacting fermionic field theory of
a new type. We study the resulting theory using renormalization group
techniques. This yields numerical estimates for the critical exponents and provides a new
alternative analytic systematic procedure to study low-dimensional directed
percolation.}\\

\end{small}
\vspace{1.5cm}
\noindent L.P.T. - ORSAY 99/05\\
\\{\small$^1$Laboratoire associ\'e au Centre National de la
Recherche Scientifique - UMR 8627}
\newpage
\section{Introduction}
\subsection{Microscopic model}\label{modelecomplet}
Each site of a one-dimensional lattice (${{\mathbb{Z}}}$) is initially  occupied by
one particle $A$ with probability $\rho$ or empty with probability
$1-\rho$. The $A$ particles perform simple (continuous time) random walk with
a diffusion constant $D$. We further impose the exclusion constraint, namely,
each site is occupied by at most one particle. In addition the particles may undergo several
reaction processes:
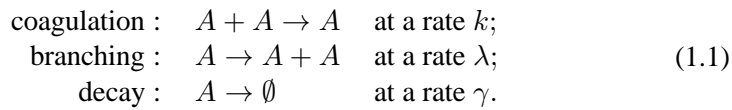
\begin{equation}\label{reac-diff}\begin{array}{rll}
\text{coagulation: }& {A+A}\rightarrow A &\text{ at a rate } k ;\\
\text{branching: }& {A}\rightarrow A+A &\text{ at a  rate } \lambda ;\\
\text{decay: }& A\rightarrow\emptyset &\text{ at a rate } \gamma.
\end{array}\end{equation}
Owing to the exclusion constraint particles react when they sit on
neighboring sites. Similarly diffusion takes place only when empty sites
allow it to. Coagulation, branching and decay, along with diffusive motion, define a reaction-diffusion process that has already received
considerable attention in the past: this is the Schl\"ogl autocatalytic
reaction, which is known to belong to the universality class of directed
percolation. The $d$-dimensional generalization of
this model has been studied via renormalization group techniques by means of $\eps$-expansion in the vicinity of the upper
critical dimension $d_c=4$ (\cite{refDPeps}). The very few analytic results that exist in one dimension are
based on short-time series expansions (\cite{refDPspin}). Our aim in this work 
is to provide a systematic approximation scheme specific to $d=1$.
\subsection{Mean field for directed percolation}
It is possible to write a mean field equation for the average local particle
density
$n(t)$ at time $t$:
\begin{equation}
\frac{\dd n}{\dd t}=(\lambda-\gamma)n-(k+\lambda)n^2
\end{equation}
From this equation one predicts that in the steady state
\begin{equation}
n(\infty)=\left\{
\begin{array}{ll}
\frac{\lambda-\gamma}{k+\lambda}&\text{ if }\lambda>\gamma\\
0&\text{ otherwise }
\end{array}\right.
\end{equation}
Hence mean field predicts a continuous transition between an active state for
$\gamma<\lambda$ in which a finite fraction of $A$'s survives indefinitely, and an
absorbing state in which $A$'s have completely disappeared forever, which occurs
for $\gamma\geq \lambda$. In the following we shall use $\gamma$ as the control parameter and fix
all other parameters. At the mean-field level we see that the steady state of this system undergoes a
second-order phase transition between an active state with nonzero $A$ density, at
$\gamma<\lambda$, and an absorbing $A$-free state at $\gamma>\lambda$. Within the mean-field
picture the transition occurs at the critical value $\gamma_c=\lambda$. It is
possible to summarize the scaling properties of the particle density in a single
formula 
\begin{equation}
n(t)=b^{-\frac{1+\eta}{2}}{\cal F}(b^{-z}t,b^{1/{\nu }}|\gamma-\gamma_c|)
\end{equation}
which holds for $b\gg 1$ with the arguments of ${\cal F}$ fixed. This scaling
relation defines the critical exponents $\eta$, $z$ and $\nu$. Their mean-field
values are $\eta=0$, $z=2$ and $\nu=1/2$. In the steady state the density
behaves as $|\gamma-\gamma_c|^\beta$ as $\gamma\rightarrow \gamma_c^-$, which defines the
exponent $\beta=\nu(1+\eta)/2$. 
\subsection{Motivations and outline}
From the analytic point of view there are very few exact or even approximate results on directed percolation in low space dimension
(see \cite{Liggett, Dickman} for recent reviews). Owing to its ubiquitous nature in the study of stochastic
processes, directed percolation has become the paradigm of out-of-equilibrium systems possessing a
second-order phase transition in their steady state. Our aim is to remedy the scarcity of analytic
techniques specific to low and physically relevant space dimensions. Indeed $d=1$ is the relevant
dimension for instance in the study of surface growth phenomena. Another motivation comes from particle
physics. There the  branching-coagulation language is used as a phenomenological description of
hadronic high-energy scattering processes with $d=2$ being the physical dimension corresponding
to the
number of transverse space
directions. Other applications include the study of intermittency (see {\it e.g.} Henkel and
Peschanski~\cite{henkelpeschanski}), either in the
Schwinger mechanism~\cite{bialas}, or in turbulence~\cite{pomeau}. Fluctuations play an increasing r\^ole as the dimension is decreased below the upper
critical dimension $d_c=4$, which
provides further motivation to focus on low dimensions. In the absence of any exact
solution we believe that our method provides new insight into the peculiarities of one-dimensional
directed percolation.\\

The outline of this paper is as follows. In section 2 we map the master
equation that describes the reaction-diffusion process
Eq.~(\ref{reac-diff}) first onto a spin-chain problem. The
spin-chain is then mapped onto a fermionic field theory. The
procedure, which we describe in great detail, consists in building a fermionic field theory starting from
a non-hermitian hamiltonian originating from the stochastic process Eq.~(\ref{reac-diff}). This raises a number of difficulties which, to our knowledge, appear for the
first time in the literature. These are exact mappings. We  then proceed with
the analysis of the full  theory describing directed percolation using renormalization group techniques.
Our calculations yield numerical estimates for the critical exponents.
\section{From the master equation to a quantum spin chain}
\subsection{Master equation}
A microstate of the system described in paragraph (\ref{modelecomplet}) is characterized by the set of occupation numbers
$\{n_j\}_{j\in\mathbb{Z}}$ defined by
\begin{equation}
n_j=\left\{
\begin{array}{ll}
1&\text{if site }j\text{ is occupied by an }A\\
0&\text{if site }j\text{ is empty}\end{array}\right.
\end{equation}
We now define the {\em spin} variable $s_j\equiv 2n_j-1$. Let $s\equiv \{s_j\}$
denote a generic microstate of the system and let it index a set of vectors
$|s\rangle$ in a Hilbert space. The master equation for the probability of
occurence $P(s,t)$ of state $s$ at time $t$ is equivalent to an evolution
equation for the linear combination
\begin{equation}
|\Phi(t)\rangle\equiv\sum_s P(s,t)|s\rangle
\end{equation}
which reads
\begin{equation}\label{Schro}
\frac{\dd |\Phi(t)\rangle}{\dd t}=-\hat{H}|\Phi(t)\rangle
\end{equation}
where $\hat{H}$ is an evolution operator (also abusively called a {\it Hamiltonian}) acting in the Hilbert space spun by the
states $|s\rangle$. We introduce the Pauli matrices ${\vec{\sigma}}_i$ defined by
\begin{equation}
\sigma_i^x=\left(\begin{array}{cc}0&1\\1&0\end{array}\right),\;\;\;
\sigma_i^y=\left(\begin{array}{cc}0&-i\\i&0\end{array}\right),\;\;\;
\sigma_i^z=\left(\begin{array}{cc}1&0\\0&-1\end{array}\right)
\end{equation}
and also define the raising and lowering operators 
\begin{equation}
\sigma_{i}^\pm\equiv\frac
12(\sigma_i^x\pm i\sigma_i^y)
\end{equation}
We thus have the identities $\sigma_i^z=2\sigma_i^+\sigma_i^--1$ and
$\sigma_i^+\sigma_i^-+\sigma_i^-\sigma_i^+=1$. The variable $s_i$ is the eigenvalue of $\sigma_i^z$. One may verify that $\hat{H}$ may be written in
the form
\begin{equation}\label{hamiltonian}
\hat{H}=\hat{H}_{\text{diffusion}}
+\hat{H}_{\text{decay}}+\hat{H}_{\text{branching}}+\hat{H}_{\text{coagulation}}
\end{equation}
where we have set
\begin{eqnarray}
\hat{H}_{\text{diffusion}}=-\frac{D}{2}\sum_i\vec{\sigma}_i.\vec{\sigma}_{i+1}\label{Hdiff}\\
\hat{H}_{\text{decay}}=\gamma\sum_i(\frac 12\sigma_i^z-\sigma_i^-)\label{Hdecay}\\
\hat{H}_{\text{branching}}=-\frac{\lambda}{4}\sum_i(\sigma_i^z\sigma_{i+1}^z
+2\sigma_i^++
\sigma_i^+\sigma_{i+1}^z+\sigma_{i+1}^+\sigma_i^z)\label{Hbranching}\\
\hat{H}_{\text{coagulation}}=\frac{k}{4}\sum_i(2\sigma_i^z-2\sigma_i^-
+\sigma_i^z\sigma_{i+1}^z-\sigma_i^-\sigma_{i+1}^z-\sigma_{i+1}^-\sigma_i^z)\label{Hcoagulation}
\end{eqnarray}
Note that we have dropped all constant terms in
Eqs.~(\ref{Hdiff}--\ref{Hcoagulation}). They ensure the conservation of probability but will however
play no r\^ole in the subsequent analysis.
\subsection{General properties of the spin chain and average of observables}
In this paragraph we recall for completeness some of the properties of the spin chain Hamiltonian defined by
Eq.~(\ref{Schro}). We need to introduce a {\it projection state} $\langle\text{\bf p}|$ defined by
\begin{equation}
\langle\text{\bf p}|\equiv\sum_s\langle s|
\end{equation}
Given a physical observable $A(s)$ we denote by $\hat{A}$ the operator obtained by replacing in the explicit
expression of $A$ the variable $s_i$ by the operator $\sigma_i^z$. For instance the choice
$A(s)=\frac 12(s_j+1)$, which is the local
number of particles at site $j$, leads to
$\hat{A}=\frac 12(\sigma_j^z+1)$. The average of the observable $A(s)$ may be expressed as
\begin{equation}
\langle A(s)\rangle(t)=\langle\text{\bf p}|\hat{A}|\Phi(t)\rangle
\end{equation}
as was first noticed by Felderhof~\cite{Felderhof}.\\

Conservation of probability imposes that $\langle\text{\bf p}|$ is a left
eigenstate of $\hat{H}$ with eigenvalue 0:
\begin{equation}
\langle\text{\bf p}|\hat{H}=0
\end{equation}
from which it follows that 
\begin{equation}
\forall t,\;\;\;\langle\text{\bf p}|\Phi(t)\rangle=1
\end{equation}
Besides, $\hat{H}$ has at least one right eigenvector with eigenvalue 0, which
describes the stationary state of the system. The eigenvalues of $\hat{H}$ all
have a positive real part. Other details may be found in the reviews by Alcaraz {\it et al.}~\cite{Malte1} or Henkel {\it
et al.}~\cite{Malte2}. For our purposes we need one more property of the
projection state. It is based on the following
identity: 
\begin{equation}\label{miracle}
\ee^{\sigma_i^-}\sigma_i^+=(\sigma^+_i-\sigma_i^--2\sigma^+_i\sigma^-_i+1)\ee^{\sigma_i^-}
\end{equation}
After noticing that
\begin{equation}
\langle \text{\bf p}|=\langle -1|\ee^{\sum_j\sigma_j^-}
\end{equation}
it becomes possible to express the average of an observable $A(s)$ in the form
\begin{equation}
\langle A(s)\rangle(t)=\langle -1| {\tilde{A}}'\ee^{-\tilde{H}t}\ee^{\sum_j \sigma_j^-}|\Phi(0)\rangle
\end{equation}
where ${\tilde{A}}'$ and $\tilde{H}$ are deduced from $\hat{A}$ and $\hat{H}$
as follows. Express these operators
only in terms of the $\sigma^+$'s, $\sigma^-$'s and $\sigma^+\sigma^-$'s. For each $j$ replace in the resulting
expression $\sigma_j^+$ by $\sigma^+_j-\sigma_j^--2\sigma^+_j\sigma^-_j+1$. This yields the operators
$\tilde{A}$ and $\tilde{H}$. In the expression of $\tilde{A}$ one first puts all the $\sigma^+$'s to the left of
the $\sigma^-$'s then one formally sets the  $\sigma^+$'s to 0. This yields the
operator ${\tilde{A}}'$. With the particular expression of $\hat{H}$
Eq.~(\ref{hamiltonian}) one finds after straightforward manipulations:
\begin{equation}\label{modifham}\begin{split}
\tilde{H}=&-(D+\frac{\lambda}{2})
\sum_i\Big[\sigma_i^-\sigma_{i+1}^++\sigma_i^+\sigma_{i+1}^--2\hat{n}_i
\Big]\\
&+(\gamma-\lambda)\sum_{i}\hat{n}_i\\
&+(\lambda+k-2D)\sum_i
\hat{n}_{i}\hat{n}_{i+1}\\
&+\frac{1}{2}(\lambda+k)\sum_i\Big[\sigma_i^-\hat{n}_{i+1}+\hat{n}_i\sigma_{i+1}^-\Big]\\
&-\frac{\lambda}{2}\sum_i\Big[\sigma_i^+\hat{n}_{i+1}+\hat{n}_i\sigma_{i+1}^+\Big]
\end{split}\end{equation}
where we have adopted the notation ${\hat{n}}_i\equiv\sigma_i^+\sigma_i^-$. As we have already mentioned we have omitted a constant term in
$\tilde{H}$. In order to find this constant, one would have to push the $\sigma^+$'s to the left of the $\sigma^-$'s then formally set to 0
all the $\sigma^+$'s, which produces an operator ${\tilde{H}}'$ and
adjust the constant so that $\langle-1|{\tilde{H}}'=0$.

\section{Fermionic field theory}
\subsection{Jordan-Wigner transformation}
We define a set of fermionic creation and annihilation operators $c_i^{\dagger}$ and $c_i$ by
\begin{equation}\label{defcc}
\sigma_i^+=c_i^{\dagger}\exp\Big[i\pi\sum_{j<i}\hat{n}_j\Big],\;\;\;\sigma_i^-=\exp\Big[-i\pi\sum_{j<i}\hat{n}_j\Big]c_i
\end{equation}
where $\hat{n}_j=\sigma_i^+\sigma_i^-=c_j^{\dagger}c_j$ is the particle number operator. It has the eigenvalue $n_j$. This is the Jordan-Wigner transformation. The
commutation relations between the $c_i$ and $c_j^{\dagger}$ are
\begin{equation}
\{c_i,c_{j}^{\dagger}\}=\delta_{ij},\;\;\{c_i^{\dagger},c_j^{\dagger}\}=0=\{c_i,c_j\}
\end{equation}
An equivalent expression for the {\it string} operator in Eq.~(\ref{defcc}) is
\begin{equation}
\text{e}^{\pm i\pi\sum_{\ell<i}\hat{n}_\ell}=\prod_{\ell<i}(1-2\hat{n}_\ell)
\end{equation}
As an example, for diffusion alone, the evolution operator would read
\begin{equation}\label{Htildediff}
\tilde{H}_{\text{diffusion}}=-
D\sum_i\Big[c_i^{\dagger}c_{i+1}+c_{i+1}^{\dagger}c_{i}-2\hat{n}_i+
2\hat{n}_i\hat{n}_{i+1}\Big]
\end{equation}
It is surprisingly left invariant by the transformation Eq.~(\ref{miracle}).
In the quadratic part of Eq.~(\ref{Htildediff}) one may recognize the diffusion process, that results in a Laplacian. However diffusion takes place with the constraint of excluded volume, which accounts for the quartic interaction term.
\subsection{Coherent states}
As we have seen in the previous section, the average of a physical observable can be cast in the form
\begin{equation}
\langle A\rangle(t)=
\langle-1|\tilde{A}'\ee^{-\tilde{H}t}\ee^{\sum_i\sigma_i^-}|\Phi(0)\rangle
\end{equation}
We shall restrict our analysis to the case of an initial state with particles
randomly and independently distributed
with local density $\rho_j$ at site $j$, that is with the initial
distribution
\begin{equation}
P(n,t=0)=\prod_j\Big[\rho_j\delta_{n_j,1}+(1-\rho_j)\delta_{n_j,0}\Big]
\end{equation}
Under those assumptions it is easy to check that
\begin{equation}
|\Phi'(0)\rangle\equiv \ee^{\sum_i\sigma_i^-}|\Phi(0)\rangle=\ee^{\sum_j\rho_j\sigma^+_j}|-1\rangle
\end{equation}
At this stage we will denote by $|0\rangle$ the vacuum ({\it i.e.} particle-free) state, instead of $|-1\rangle$, as the former notation is more appropriate in the particle number language. To summarize, the average of an observable $A$ reads
\begin{equation}\label{solution}
\langle A\rangle(t)=\langle 0|\tilde{A}' \ee^{-\tilde{H}t}|\Phi'(0)\rangle
\end{equation}
We rewrite the exponential factor in Eq.~(\ref{solution}) using Trotter
formula:
\begin{equation}
\text{e}^{-\tilde{H}t}=\lim_{N\rightarrow\infty}\Big(1-\frac{\tilde{H}t}{N}\Big)^{N}
\end{equation}
We now introduce the coherent states associated to the creation and
annihilation operators. Let $|\psi_{ij}\rangle$, for $j=0,...,N$ and
$i\in{\mathbb{Z}}$
denote a set of coherent 
states associated to $c_i$ and $c_i^{\dagger}$:
\begin{eqnarray}\label{defcoh}
|\psi_{ij}\rangle=
\text{e}^{-\frac{1}{2}\psi^*_{ij}\psi_{ij}}
\text{e}^{-\psi_{ij}c_i^{\dagger}}|0\rangle\nonumber\\
\langle\psi_{ij}|=
\langle 0|\text{e}^{-\frac 12 \psi_{ij}^*\psi_{ij}}
\text{e}^{-c_i\psi_{ij}^*}
\end{eqnarray}
These coherent  states are indexed by a pair of conjugate Grassmann
variables $\psi_{ij}$ and $\psi_{ij}^*$. By definition they have the property
that
\begin{equation}
c_i|\psi_{ij}\rangle=\psi_{ij}|\psi_{ij}\rangle,\;\;\;\langle\psi_{ij}|c_i^{\dagger}=\langle\psi_{ij}|\psi_{ij}^*
\end{equation} 
With the definition
Eq.~(\ref{defcoh}) one may verify that
\begin{equation}
\langle\psi_{ij}|\psi_{ij}\rangle=1
\end{equation}
and that, with the integration convention
$\int\dd\psi\;\psi=\int\dd{\psi}^*\;\psi^*=1$, one has
\begin{equation}
\int\dd\psi_{ij}^*\dd\psi_{ij}\text{e}^{-\psi_{ij}^*\psi_{ij}}\psi_{ij}^n
\psi_{ij}^{*m}=\delta_{nm}\text{\bf}
\end{equation}
which implies
\begin{equation}\label{completude}
\int
\dd\psi_{ij}^*\dd\psi_{ij}|\psi_{ij}\rangle\langle\psi_{ij}|=\text{\bf
1}
\end{equation}
Eq.~(\ref{completude}) generalized to the state
\begin{equation}
|\psi_j\rangle\equiv\otimes_i |\psi_{ij}\rangle
\end{equation}
leads us to 
\begin{equation}\begin{split}
\text{e}^{-\tilde{H}t}=\int\prod_{j=0}^N \prod_i
\Big[\dd\psi_{ij}^*\dd\psi_{ij}\Big]|\psi_{j=N}\rangle&\langle\psi_{N}|
1-\tilde{H}\frac{t}{N}|\psi_{N-1}\rangle...\\
&\langle\psi_{1}|1-\tilde{H}\frac{t}{N}
|\psi_{0}\rangle\langle\psi_{0}|
\end{split}\end{equation}
Thus we need the quantities
\begin{equation}
H_{j,j-1}\equiv\frac{\langle\psi_{j}|
\tilde{H}|\psi_{j-1}\rangle}{\langle\psi_{j}|\psi_{j-1}\rangle},\;\;\;j=1,...,N
\end{equation}
as well as
\begin{equation}
\langle\psi_{j}|\psi_{j-1}\rangle=\text{e}^{-\sum_i\psi_{ij}^*(\psi_{ij}-\psi_{i,j-1})}\text{e}^{\frac
12 \sum_i\psi_{ij}^*\psi_{ij}-\frac 12\sum_i \psi_{i,j-1}^*\psi_{i,j-1}}
\end{equation}
In order to obtain $H_{j,j-1}$ one replaces in the normal ordered form of
$\tilde{H}$ the annihilators $c_\ell$ by $\psi_{\ell,j-1}$ and the creators
$c_m^{\dagger}$ by $\psi^*_{mj}$.
The projection onto the initial state $|\Phi'(0)\rangle$ reads
\begin{equation}\begin{split}
\langle\psi_{j=0}|\Phi'(0)\rangle
=...
\text{e}^{\rho_i\psi_{i0}^*}\text{e}^{\rho_{i+1}\psi_{i+1,0}^*}...\equiv
P_i[\psi^*(x,0)]
\end{split}\end{equation}
In a similar way, one finds
\begin{equation}
\langle 0|\tilde{A}'|\psi_{j=N}\rangle=A'(\psi_N)
\end{equation}
where $A'$ is some function of $\psi_N$ the precise determination of which we postpone to paragraph \ref{expect}.
\subsection{Continuous field theory}
It is convenient to split $H_{j+1,j}$ in the form
\begin{equation}
H_{j+1,j}=\textsf{E}_{j+1,j}+{\textsf{O}}_{j+1,j}
\end{equation}
where $\textsf{E}_{j+1,j}$ and $\textsf{O}_{j+1,j}$ denote the even and odd components of $H_{j+1,j}$,
respectively. In the limit $\Delta t\rightarrow 0$ the ordered product
\begin{equation}
\overleftarrow{\Pi}\equiv (1-\Delta t H_{N,N-1})...(1-\Delta t H_{1,0})
\end{equation}
does not reduce to a simple exponential as the terms in the product do not
commute. We may use an identity (\cite{DDM}) for a sequence of odd Grassmann variables
$\alpha_1$,..., $\alpha_N$:
\begin{equation}\label{mandonnet}
\ee^{\alpha_N}...\ee^{\alpha_1}=
\ee^{\sum_n\alpha_n+\sum_{n=1}^N\sum_{m=1}^{n-1}\alpha_n\alpha_m}
\end{equation}
Hence we may rewrite $\overleftarrow{\Pi}$ in the form
\begin{equation}
\overleftarrow{\Pi}=\exp\Big[-\Delta
t\sum_{j}H_{j+1,j}+\Delta t^2\sum_{j=0}^{N-1}\sum_{\ell=0}^{j-1}
\textsf{O}_{j+1,j}\textsf{O}_{\ell+1,\ell}\Big]
\end{equation}
which yields in the continuum limit:
\begin{equation}\begin{split}
\overleftarrow{\Pi}=&\exp\Big[-\int_0^{t_f}\dd t H[\psi(t),\psi^*(t)]\\
&+\int_0^{t_f}\dd t\dd
t'\Theta(t-t')\textsf{O}[\psi(t),\psi^*(t)] \textsf{O}[\psi(t'),\psi^*(t')]\Big]
\end{split}\end{equation}
We define an action $S$ by
\begin{equation}\begin{split}
S[\bar{\psi},\psi]=&\int \dd x\;\dd
t\;\bar{\psi}\p_t\psi\\&+\int\dd t\;H[\psi,\bar{\psi}]\\
&-\int\dd t\;\dd t'\;\Theta(t-t')\textsf{O}[\psi(t),\bar{\psi}(t)] 
\textsf{O}[\psi(t'),\bar{\psi}(t')]
\end{split}\end{equation}
where the notation $H[\psi(t),\bar{\psi}(t)]$ means the (normal-ordered)
$\tilde{H}$ in which the creators
and annihilators have been replaced by their Grassmann eigenvalues after the
former have been moved to the left of the latter. The quantity
$\textsf{O}[\psi(t),\bar{\psi}(t)]$ is the odd component of
$H[\psi(t),\bar{\psi}(t)]$. Thus we come up with an action that comprises terms
that are nonlocal in space and time. In what follows it could be possible to proceed
entirely within the fermion operator formalism (pertubation expansions
have been checked this way). There one would have to
use time-ordered products in evaluating correlation functions, the effect
of which is precisely to rebuild effectively nonlocal interactions.\\

Again using Eq.~(\ref{mandonnet}), we find
\begin{equation}
P_i[\bar{\psi}(x,0)]=\exp\Big[\sum_i
\rho_i\bar{\psi}_i(0)+\sum_{i<j}\rho_i\rho_j\bar{\psi}_i(0)\bar{\psi}_j(0)\Big]
\end{equation}
which allows us to express the average of the physical observable $A$ in a path
integral form as follows:
\begin{equation}
\langle A\rangle(t_f)=\int {\cal D}\bar{\psi}{\cal D}\psi
A'[\psi(x,t_f)]\ee^{-S[\bar{\psi},\psi]}P_i[\bar{\psi}(x,0)]
\end{equation}
We are now in a position to apply the existing techniques of field theory.
\subsection{Comparison with bosonic field theories for stochastic processes}
To conclude this paragraph we would like to make a comparison with reaction-diffusion
problems in which the exclusion constraint is not imposed
(\cite{Peliti,Cardy2}). There bosonic
operators ($a_i,a_i^\dagger$), rather than spin operators, are used to express the
evolution operator $\hat{H}$. The projection state reads
\begin{equation}
\langle\text{\bf p}|=\langle 0|\ee^{\sum_i a_i}
\end{equation}
and the bosonic counterpart to identity Eq.~(\ref{miracle}) is
$\ee^{a_i}{a_i^\dagger}^n=(a^\dagger_i+1)^n\ee^{a_i}$. The commutation of the
factor $\ee^{\sum_i a_i}$ through the $\ee^{-\hat{H}t}$ can certainly be
performed from the outset, but is not compulsory. It is possible by an
appropriate shift of the field associated to the creation operators to perform
this transformation directly within the functional integral formulation. In the
present case it seems that such an {\it a posteriori} nonlinear change of Grassmann
field having the same effect as the nontrivial transformation
Eq.~(\ref{miracle}) would be difficult to find and to handle.

\subsection{Expectation values of some observables}\label{expect}
The main observable of interest is the local particle number $A=n_i$. Hence $\hat{A}=\hat{n}_i$ and $\tilde{A}=\hat{n}_i+\sigma_i^-$, so that $\tilde{A}'=\sigma_i^-$. Since
\begin{equation}
\langle 0|\tilde{A}'=\langle 0|c_i 
\end{equation}
we can write the average of the local particle number at site $i$ as the
functional integral
\begin{equation}
\langle n_i\rangle(t)=\int {\cal D}\bar{\psi}{\cal D}\psi\;
\psi_i(t)\;\text{e}^{-S[\psi,\bar{\psi}]}P_i[\bar{\psi}]
\end{equation}

Another physical quantity of interest is the equal time two-point correlation function $A=n_i n_j$, with $i<j$. The associated $\tilde{A}'$ reads
\begin{equation}
\tilde{A}'=\sigma_i^-\sigma_j^-=\prod_{\ell=i}^{j-1}(1-2\hat{n}_\ell)c_jc_i
\end{equation}
hence $\langle 0|\tilde{A}'=\langle 0|c_j c_i$ which yields the equal time correlation function
\begin{equation}
\langle n_in_j\rangle(t)=\int {\cal D}\bar{\psi}{\cal D}\psi\;
\psi_j(t)\psi_i(t)\;\text{e}^{-S[\psi,\bar{\psi}]}P_i
\end{equation}
For $i=j$ the above formula does not hold, and one finds $\tilde{A}'=c_i$ so that one recovers $\langle n_i^2\rangle=\langle n_i\rangle$. Other equal time averages can be derived in a similar fashion.\\




\section{Directed Percolation}
\subsection{Action}
We now return to our problem of interest, directed percolation. It is described
by the evolution operator Eq.~(\ref{modifham}), which splits into several
contributions. We find it convenient to introduce the non local functional 
\begin{equation}
\xi_i(t)\equiv\prod_{\ell\leq i}\text{e}^{-2\bar{\psi}_\ell(t)\psi_\ell(t)}
\end{equation}
which takes into account the string operator present in the definition of the
Jordan-Wigner fermions. The action describing the whole reaction-diffusion process reads
\begin{equation}\begin{split}
S_{\text{DP}}=&\sum_i\int\dd
t\Bigg[\bar{\psi}_i\left(\p_t+\gamma-\lambda-(D+\frac{\lambda}{2})\Delta\right)\psi_i
\\&+(\lambda+k-2D)\bar{\psi}_i\psi_i\bar{\psi}_{i+1}\psi_{i+1}\\&
-(\lambda+k)\bar{\psi}_i\psi_i\xi_{i-1}\p\psi_i\\
&+\lambda\bar{\psi}_i\psi_i\xi_{i-1}\p\bar{\psi}_i\Bigg]\\&
-\sum_{i,j}\int\dd t\dd t'\Theta(t-t')
\left[-(\lambda+k)\bar{\psi}_i\psi_i\xi_{i-1}\p\psi_i
+\lambda\bar{\psi}_i\psi_i\xi_{i-1}\p\bar{\psi}_i\right](t)\times\\&
\left[-(\lambda+k)\bar{\psi}_j\psi_j\xi_{j-1}\p\psi_j
+\lambda\bar{\psi}_j\psi_j\xi_{j-1}\p\bar{\psi}_j\right](t')
\end{split}
\end{equation}
with $\p f_i\equiv\frac{1}{2}(f_{i+1}-f_{i-1})$ and $\Delta f_i\equiv
f_{i+1}-2f_i+f_{i-1}$. Finally note that we have omitted terms (located
at the time slice $t=0$) describing the
initial state of the system.
\subsection{Scaling analysis}
After rescaling fields and coupling constants we may
take 
\begin{equation}\label{actionfinale}\begin{split}
S[\bar{\psi},\psi]=&\int\dd
x\dd t\;\left[\bar{\psi}(\p_t+D\sigma-D\p_x^2))\psi+Dg\psi\bar{\psi}\xi
(\p_x\psi-\p_x\bar{\psi})\right]\\
&+\int\dd x\dd y\dd t\dd t'
\left[u_1\{\psi\bar{\psi}\xi\p_x\psi\}_{(x,t)}
\Theta(t-t')\{\psi\bar{\psi}\xi
\p_y\bar{\psi}\}_{(y,t')}\right.
\\&+
u_2\{\psi\bar{\psi}\xi\p_x\bar{\psi}\}_{(x,t)}
\Theta(t-t')\{\psi\bar{\psi}\xi
\p_y{\psi}\}_{(y,t')}\\&+
u_3\{\psi\bar{\psi}\xi\p_x\psi\}_{(x,t)}
\Theta(t-t')\{\psi\bar{\psi}\xi
\p_y{\psi}\}_{(y,t')}\\&\left.+
u_4\{\psi\bar{\psi}\xi\p_x\bar{\psi}\}_{(x,t)}
\Theta(t-t')\{\psi\bar{\psi}\xi
\p_y\bar{\psi}\}_{(y,t')} \right]
\end{split}\end{equation}
as the starting point of the subsequent analysis. The coupling constants are
$u_1=u_2=-u_3=-u_4=(Dg)^2$. We find the field
and coupling constants dimensions by power counting in momentum unit:
\begin{equation}
[\psi]=[\bar{\psi}]=\frac{1}{2},\;\;[x^2]=[t]=-2,\;\;[g^2]=1
\end{equation}
There is an interesting difference with the one-dimensional Reggeon field
theory: there the equivalent coupling
constant $g^2$ has dimension $[g^2]=\eps=4-d=3$. It is natural to expect, then,
that perturbation will lead to more sensible results than the mere $d\rightarrow 1$
extrapolation of the $\eps$-expansion.\\

In the action Eq.~(\ref{actionfinale})
the nonlocal exponential factors 
\begin{equation}\xi(x,t)=\exp(-2\int_{-\infty}^x\dd
y\bar{\psi}\psi(y,t))
\end{equation}
play a crucial r\^ole. Indeed, by construction their presence breaks the space
$x\to-x$ symmetry of the action, along with translation invariance. This signals
that the right quantities to study in a renormalization perpective are not the
usual vertex functions.\\

Finally note the absence of the quartic interaction term
of the form 
\begin{equation}
\int\dd t\sum_i\bar{\psi}_i\psi_i\bar{\psi}_{i+1}\psi_{i+1}
\end{equation}
It may
readily be seen that this term can be written in the form
\begin{equation}\begin{split}
\int\dd t\sum_i\bar{\psi}_i\psi_i\bar{\psi}_{i+1}\psi_{i+1}=
&\int\dd t\frac{\dd q_1}{2\pi}\frac{\dd q_2}{2\pi}\frac{\dd q_3}{2\pi}\frac{\dd
q_4}{2\pi}\\&\times
(2\pi)\delta(q_1+q_2+q_3+q_4)(\ee^{i(q_3+q_4)}-1-i(q_3+q_4))\\
&\times\bar{\psi}(q_1,t)\psi(q_2,t)\bar{\psi}(q_3,t)\psi(q_4,t)
\end{split}
\end{equation}
which, in the small momentum region yields an irrelevant operator of {\it negative}
dimension equal to -1. 
\subsection{Correlation functions and renormalization}
At this stage we have to choose a renormalization scheme. We have explained why
the natural choice of
examining the divergences of the vertex functions is not adapted here since we
are dealing with a theory that neither has translation invariance nor space
reflection symmetry. This is due to the anisotropic construction of the
fermion operators. In fact the two-point vertex function can still be used to
renormalize the fields, the diffusion constant and the mass. As for the
coupling constant we have chosen a particular class of correlation functions in
which space arguments are carefully ordered. As could be expected the perturbation series organizes in powers of the
dimensionless coupling $g^2/\sqrt{\sigma}$, and our interest lies in the critical regime
$\sigma\rightarrow 0$. Renormalization is thus needed to extract physical
information from the above naive expansion.\\

\begin{figure}
$$\input{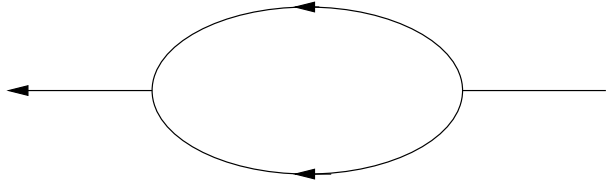}$$
\caption{A plain leg denotes a $\psi$ field while an
arrowed one denotes a conjugate $\bar{\psi}$. This is a diagram contributing to
the renormalization of the propagator. Two of the four internal legs carry a
derivative. In Fourier space, the loop integral has to take into account the
product of the momenta carried by those legs.}\label{figure1}
\end{figure}

\begin{figure}
$$\input{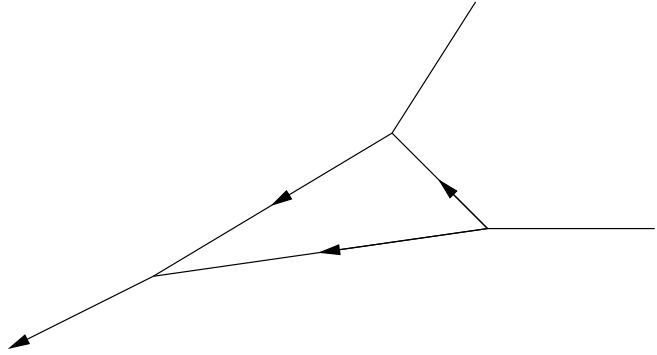}$$
\caption{One of the diagrams entering the renormalization of $g$. It involves
$u_1$ and $g$.}\label{figure2}
\end{figure}

We will renormalize the massive theory at zero external frequency and momentum
(\cite{Parisi}). 
We define the
renormalized fields and parameters by
\begin{eqnarray}
\psi=
Z^{1/2}_\psi\psi_{\sR}\nonumber\\
\bar{\psi}=
Z^{1/2}_\psi\bar{\psi}_{\sR}\nonumber\\
D=Z_{\psi}^{-1}Z_D D_{\sR}\nonumber\\
(\bar{\psi}\psi)=Z^{-1}_\sigma(\bar{\psi}\psi)_{\sR}\nonumber\\
g\equiv Z_{D}^{-1}Z_\psi^{-1/2}Z_g g_{\sR}\sigma_{\sR}^{1/4}
\end{eqnarray}
The first three $Z$-factors are determined by the following conditions:
\begin{eqnarray}
\frac{\p\Gamma_{\sR}^{(1,1)}}{\p(-i\omega)}(0,0)=1,\;\;
\frac{\p\Gamma_{\sR}^{(1,1)}}{\p
k^2}(0,0)=D_{\sR},\;\;\Gamma^{(1,1)}_{\sR}(0,0)=D_{\sR}\sigma_{\sR}
\end{eqnarray}
The one-loop expression for the propagator, calculated after the diagram
depicted in Fig.\,\ref{figure1}, reads
\begin{equation}\begin{split}
\Gamma^{(1,1)}(k,\omega)=&-i\omega\left(1-\frac{g^2}{2}\int\frac{\dd
q}{2\pi}\frac{q^2}{(q^2+\sigma)^2}\right)\\
&+Dk^2\left(1-g^2\int\frac{\dd
q}{2\pi}\frac{q^2\sigma}{(q^2+\sigma)^3}\right)\\
&+D\sigma+g^2 \int\frac{\dd
q}{2\pi}\frac{q^2}{q^2+\sigma}
\end{split}\end{equation}
An explicit computation of the integrals yields
\begin{equation}\begin{split}
\Gamma^{(1,1)}(k,\omega)=&-i\omega\left(1-\frac{g^2}{8\sqrt{\sigma}}\right)
+Dk^2\left(1-\frac{g^2}{16\sqrt{\sigma}}\right)\\
&+D\sigma+g^2 \int\frac{\dd
q}{2\pi}\frac{q^2}{q^2+\sigma}
\end{split}\end{equation}
In order to renormalize the composite operator $\bar{\psi}\psi$ one notes, as in
\cite{Cardy1} that a vertex function with a $\bar{\psi}\psi$ insertion is obtained by
differentiating the corresponding vertex function with respect to $D\sigma$ and we choose the
condition $\Gamma^{(1,1;1)}_{\sR}(0,0)=1$. As for the renormalization of the coupling constant $g$ we may choose to look at
the correlation function
\begin{equation}
G^{(1,2)}(x_1,t_1;x_2,t_2,x_3,t_3)\equiv\langle\bar{\psi}(x_1,t_1)\psi(x_2,t_2)
\psi(x_3,t_3)\rangle
\end{equation}
with the restriction that $x_2<x_1<x_3$. Following the
physical explanation proposed by Cardy~\cite{Cardy1}, one understands it is more
convenient to look at the the space integral of $G^{(1,2)}$ with the space
arguments kept ordered as specified. We
now define a quantity very similar to a vertex function: we first continue the latter function correlation function in which the arguments are
ordered to unconstrained space arguments. This does allow for an easy
calculation of its Fourier transform. We define the associated 1PI function
as $\Gamma^{(1,2)}(k_1,k_2,k_3)$ (by convention, leg 3 carries the
derivative $i k_3$). We emphasize that {\it this is not} the usual vertex function as
defined by the Legendre transform of the logarithm of the partition
function. We define the renormalized coupling $g_{\sR}$ by
\begin{equation}
\frac{\p\Gamma_{\sR}^{(1,2)}}{\p(ik_3)}(\{0,0\})=2D_{\sR} g_{\sR}
\end{equation}
We refer the reader to Fig.\,\ref{figure2} for the corresponding one-loop
diagram. A rather tedious calculation yields
\begin{equation}\begin{split}
\frac{\p\Gamma^{(1,2)}}{\p(ik_3)}(\{0,0\})&=2D g\left(1-\frac{g^2}{2}\int\frac{\dd
q}{2\pi}\frac{q^4}{(q^2+\sigma)^3}+\frac{5g^2}{4}\int\frac{\dd
q}{2\pi}\frac{q^2}{(q^2+\sigma)^2}\right)\\&=2Dg\left[1-\frac{7}{32}\frac{g^2}{\sqrt{\sigma}}\right]
\end{split}\end{equation}
so that the $Z$-factors read
\begin{equation}
Z_\psi=1+\frac{g^2_{\sR}}{8},\;\;Z_D=1+\frac{g^2_{\sR}}{16},\;\;
Z_\sigma=1+\frac{g^2_{\sR}}{4},
\;\;Z_g=1+\frac{7g_{\sR}^2}{32}
\end{equation}
Denoting the renormalized correlation length by $\xi\equiv \sigma_{\sR}^{-1/2}$, the
one-loop Wilson functions $\gamma_i\equiv-\xi\frac{\dd\ln
Z_i}{\dd\xi}$,
$i=\psi,D,\sigma,g$ follow:
\begin{equation}
\gamma_\psi\simeq -\frac{g_{\sR}^2}{8},\;\;\gamma_D\simeq-\frac{g^2_{\sR}}{16},
\;\;\gamma_\sigma\simeq-\frac{g^2_{\sR}}{4}
,\;\;\gamma_g\simeq -\frac{7 g_{\sR}^2}{32}
\end{equation}
The $\beta$-function reads
\begin{equation}
\beta_g\equiv -\xi\frac{\dd
g_{\sR}}{\dd\xi}=
g_{\sR}\left(-\frac{1}{2}+\gamma_D+\frac{1}{2}\gamma_\psi-\gamma_g\right)
\end{equation}
hence, to one-loop,
\begin{equation}
\beta_g=-g_{\sR}\left(\frac{1}{2}-\frac{3}{32}g^2_{\sR}\right)
\end{equation}
The renormalization group flow then has a single stable fixed point
$g^{*2}_{\sR}=\frac{16}{3}$. The critical exponents can be expressed in terms of
the Wilson functions at the fixed point:
\begin{eqnarray}
\eta=\gamma_\psi^*\simeq -\frac{2}{3}\nonumber\\
z=2-\gamma_D^*+\gamma^*_\psi\simeq \frac{5}{3}\nonumber\\
\nu^{-1}=2-\gamma_D^*+\gamma_\sigma^*\simeq \frac{4}{3}
\end{eqnarray}
The above numerical values are in reasonable agreement with the precise
numerical estimates of Dickman ~\cite{Dickman2} ($\eta=-0.405$, $z=1.767$,
$\nu=1.142$).
\section{Conclusions}
\subsection{Comparison with existing approaches}
Of course the method we have presented suffers two obvious drawbacks. First the
numerical values of the
critical exponents poorly compare to those given by numerical simulations. This is due to
the
absence of a small parameter validating the first order loop expansion. From that point
of view the situation
is quite similar to the low-dimensional extrapolation of $\eps$-expansions. A related
question concerns the
convergence of the series expansion of the exponents in terms of the fixed-point
renormalized coupling. Second we lack a decisive argument in favor of the
renormalizability of the theory, a property we trust in, but that we were unable to
establish on a rigorous basis. Arguments in favor of the existence of
consistent renormalization rely on the locality of the original spin
theory. The scheme that we have introduced, based upon the study of space
ordered correlation functions (and their related loop integrals), must now
be extended to all orders. 
However we emphasize that we have built an entirely  new and systematic analytic technique that
provides to our
knowledge the only
alternative to $\eps$-expansion of the Reggeon field theory and to short time series
expansions. The
approximation scheme we have developped has in principle a wide range of applicability and exhibits
many analogies with its bosonic counterpart. Its main advantage is that it naturally
incorporates the
simplification induced by the one-dimensional topology. As a final comment, we find it
amazing enough that the present
field theory has appeared in a
slightly different form in a quite different context~\cite{Cardy1}.

\subsection{Prospects}
We have also applied the above formalism (\cite{nosotros}) to the much studied
annihilation reaction
$A+A\rightarrow \emptyset$. The methods of integrable
spin chains (\cite{Malte1,Malte2}) work at a particular diffusion
contant-annihilation rate ratio (which corresponds to a free
fermion Hamiltonian Eq.~(\ref{hamiltonian})) and leads, as expected, to a $1/\sqrt{t}$ decay of the
particle density. The present method very easily allows to extend this result to any value of the
parameters.\\

There exists another class of reaction-diffusion problems, namely branching and annihilating
random walks (\cite{CardyTauber}), in which the formalism we
have developped should profitably apply. We expect it to provide a better qualitative picture
in one dimension than previous attempts (\cite{CardyTauber}) using bosonic theory down from
the upper critical dimension. It also remains to see how the formalism can be extended to
cope with several species reaction-diffusion problems ({\it e.g.}
\cite{Schmittmann}) for which no analytical nor numerical results exist
in one dimension. Our tool should prove useful in the study of
those problems, where other techniques have, up to now, failed.\\

\noindent {\bf Acknowledgments:} FvW and VB would like to thank P. Degiovanni, M. Henkel, H.J. Hilhorst, T. Jolicoeur and A. M\'ezard for the many discussions that have led to the present
work. FvW would like to thank J. Cardy for several discussions on the connections between
the present work and \cite{Cardy1}. KO would like to acknowledge financial support of the
DFG (Sonderforschungsbereich 237) and discussions with
H.K. Janssen.

\newpage
$$\text{\large{\bf FIGURES}}$$

\newpage

\end{document}